\documentclass[letterpaper, aps, showpacs,onecolumn]{revtex4}
\usepackage{amsmath}
\usepackage{amssymb}
\usepackage{latexsym}
\usepackage{graphicx,epstopdf,ifpdf,graphics}
\usepackage{hyperref}
\include{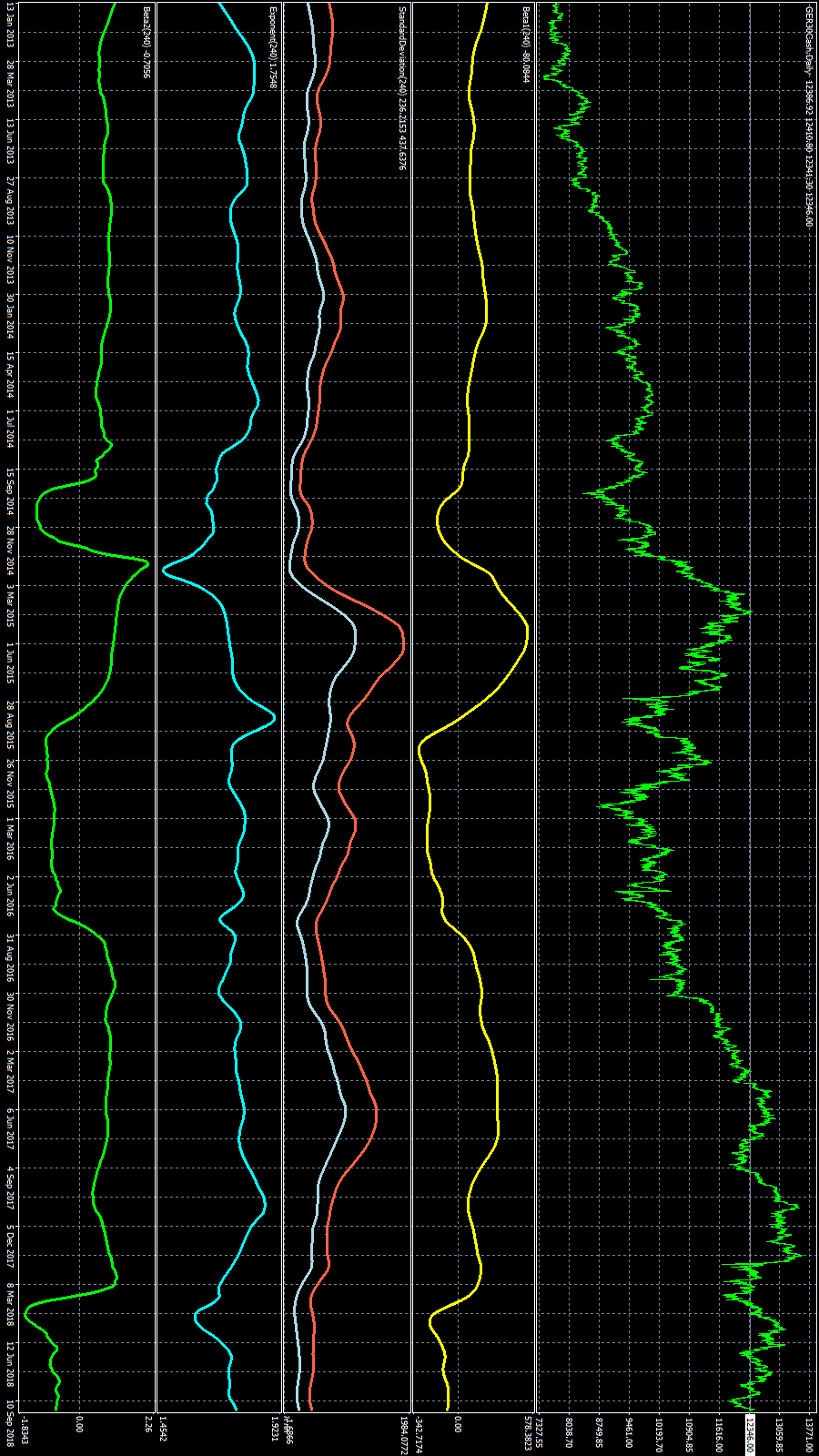}
\include{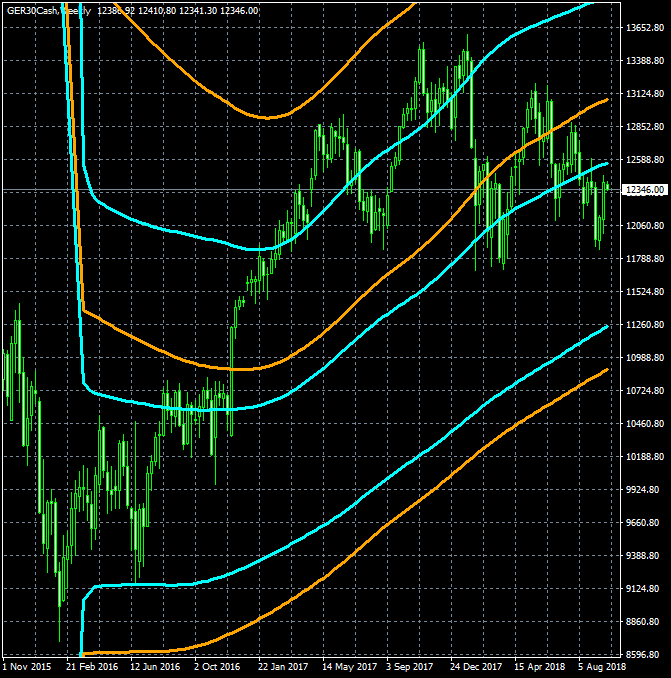}
\include{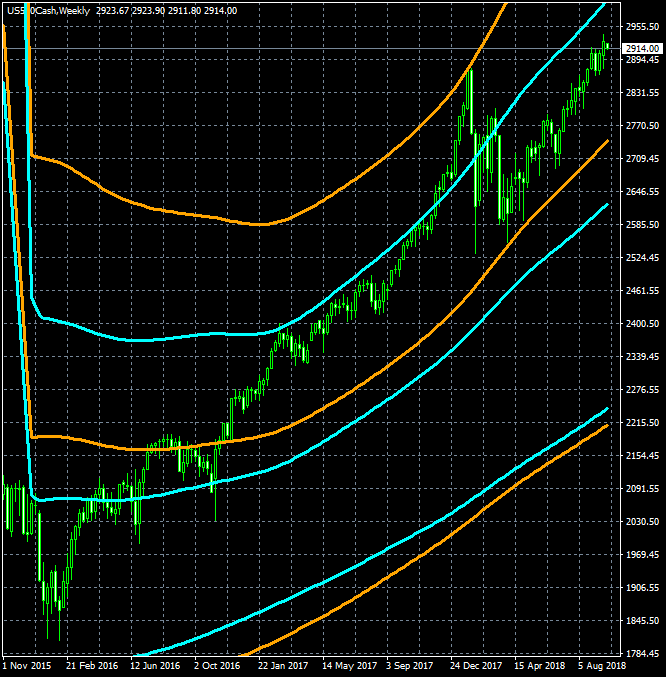}
\include{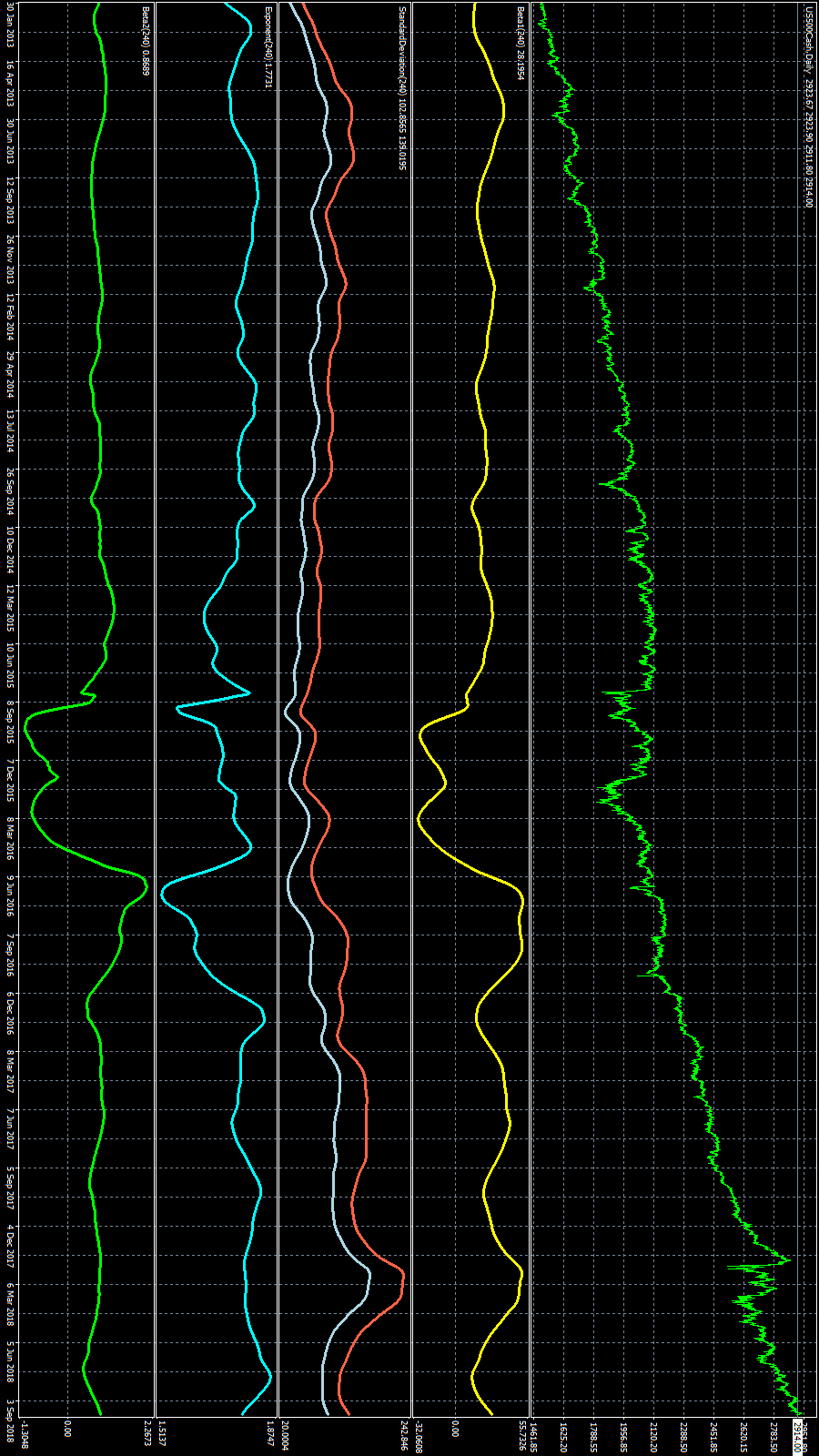}

\begin{document}
%\begin{frontmatter}
%\numberwithin{equation}{section}
%\pagestyle{fancy} 

\title{Statistical mechanics and time-series analysis by L\'evy-parameters\\
with the possibility of real-time application}

\author{Alexander Jurisch}

\affiliation{ajurisch@ymail.com, Munich, Germany}

\begin{abstract}
We develop a method that relates the truncated cumulant-function of the fourth order with the L\'evian cumulant-function. This gives us explicit formulas for the L\'evy-parameters, which allow a real-time analysis of the state of a random-motion. Cumbersome procedures like maximum-likelihood or least-square methods are unnecessary. Furthermore, we treat the L\'evy-system in terms of statistical mechanics and work out it's thermodynamic properties. This also includes a discussion of the fractal nature of relativistic corrections. As examples for a time-series analysis, we apply our results on the time-series of the German DAX and the American S\&P-500\,.
\end{abstract}

\maketitle

%\end{frontmatter}
%\tableofcontents
%*****************************************************************************
%\begin{mainmatter}

\section{Introduction}
The statistics of a macroscopic many-particle system with negligible interaction is usually well-described by a Gaussian, that may be given by the Maxwell-Boltzmann distribution or a Ginzburg-Landau functional without non-linear terms. However, there are macroscopic systems where this is not the case. This may be due to interactions, or just due to the fact that the particle-number, or the number of players if socio-economic systems are of concern, is not large enough to get even approximately close to the domain of the central-limit theorem, which is equivalent to the Gaussian. In such systems higher order fluctuations like the skewness or the kurtosis have to be taken into account, which may lead to a significant deviation from the Gaussian. 

A natural generalization of the Gaussian is given by the L\'evy-distribution. Note that a multitude of distribution-functions are special cases of the L\'evy-distribution, see e.g. the textbooks by Feller \cite{Feller} or Zolotarev \cite{Zolotarev}. A different type of generalization is provided by the Tsallis-distribution. However, the Tsallis-distribution gives rise to a non-extensive thermodynamics, and we prefer to keep in touch with the property of extensiveness. As examples for non-Gaussian macroscopic systems we chose stock-market data. However, our results are general and thus will apply to any other macroscopic system, that does not belong to the domain of the central-limit theorem.

So why stock-markets as an example for application? Because their properties with respect to non-Gaussian behaviour are quite well understood, such that they are a perfect means to test the fidelity of our approach. We have written this paper in a way that a reader who is unfamiliar with stock-markets can still take away the general essentials - or at least we hope so.

In 1961 Mandelbrot \cite{Mandelbrot} has shown that the behaviour of markets is essentially non-Gaussian. Since the first analysis of market-behaviour by Bachelier \cite{Bachelier} in 1900\,, Mandelbrot's work can be regarded as a major breakthrough. In particular, Mandelbrot has shown that the distribution-function of the random-motion of markets behaves like a L\'evy-distribution, which we shall call a L\'evian in the following. Mandelbrot's results have been confirmed by e.g. Mantegna et. al. \cite{Mantegna}. The non-Gaussian behaviour of markets is also reflected by the fact that the Gaussian Black-Scholes-Merton theory of option-pricing \cite{Black, Merton} works only insufficiently when the historic volatility (variance) is used, such that the concept of an implied volatility had to be introduced. Moreover, approaches that rely on the Cox-Ingersoll-Ross model \cite{Cox}, e.g. Tompkins \cite{Tompkins}, show a much better success in the description of option-pricing, and thus the description of the market as a whole.

However, the multitude of different models, we shall also mention Heston \cite{Heston} as a well-known one, does not make things easier, and thus it seems to be promising to walk on the general path of fractional diffusion. For an exhaustive introduction into fractional or anomalous diffusion see the report by Metzler et. al. \cite{Metzler}. Major advances in the fractional analysis of markets have been achieved e.g. by Cont et. al. \cite{Cont}, where it has been shown that markets in general show a behaviour that is related to a L\'evy-exponent $\alpha_{2}=1.7$\,. The Gaussian case would mean $\alpha_{2}=2$\,. Extensive work on the fractional behaviour of markets has been carried out by e.g. Bouchaud et. al. \cite{Bouchaud}, while e.g. Feigenbaum et. al. \cite{Feigenbaum} and e.g. Johansen et. al. \cite{Johansen1, Johansen2} focus on the critical dynamics of markets close to crashes. In the language of physics, crashes can be understood as phase-transitions. A major result of this paper is that the L\'evy-exponent $\alpha_{2}$ is a function of time or the temperature. Since $\alpha_{2}$ describes the transport-properties of a stochastic system, a continuous variation of $\alpha_{2}$ may very well be interpreted as a continuous phase-transition. Below we shall give an exhaustive discussion about the nature of the phase-transitions that may occur, and about the statistical mechanics and thermodynamics of a L\'evian system. Particularly, we shall discuss the relationship between relativistic corrections to the classical kinetic energy and fractal statistics.

Recent approaches to fractional option-pricing have been suggested by e.g. Aguilar et. al. \cite{Aguilar1, Aguilar2}, Borland et. al. \cite{Borland1, Borland2}, Kleinert et. al. \cite{ Kleinert1, Kleinert2}, and ourselves \cite{Jurisch}. Aguilar et. al. use a very technical approach that relies completely on the L\'evian, Borland et. al. suggest a model that introduces fractional behaviour by the Tsallis-distribution. More closely related to the L\'evian are the ans\"atze by Kleinert et. al. and ourselves. Kleinert et. al. introduce fractional behaviour by the use of fractional derivatives, while we suggest the use of a fractional Gaussian that is directly derived from the L\'evy-Khintchine theorem and an extremal principle, closely related to the L\'evian.

However, all models need input in terms of the numerical values of the parameters they rely on. The methods of choice to determine parameters of distribution-functions are the frequentist approaches of the least squares and the maximum-likelihood principle, or Bayesian methods. In terms of the L\'evian fitting-procedures are particularly difficult, since the L\'evian is defined by it's Fourier-transform and will only be stable for certain numerical values of the parameters it depends on. This holds for both, the normal L\'evian and the truncated L\'evian.

In this paper we develop a theory, that allows us to calculate the parameters of the L\'evian directly from the time-series under consideration, but without the need to employ frequentist or Bayesian fitting-procedures. This, in fact, allows us to calculate the L\'evy-parameters in real-time, such that we can make statements about the local state a random-motion dwells in. In detail, our approach relates the truncated cumulant-function of the fourth order with the cumulant-function of the L\'evian by a continuous matching with logarithmic conditions. The truncated cumulant-function is but polynomial, and thus does not yield a stable result. Our way to solve this problem is to perform a Pad\'e-resummation on the truncated cumulant-function, from which a stable result can be obtained. We apply our results on the random-motions of the German DAX and the American S\&P 500 in terms of a state-analysis from 2013 to 2018\,. Our results are in good agreement with the behaviour of the indices. Especially we emphasize, that our local results for the L\'evy-exponent $\alpha_{2}$ as a function of time are in very good agreement with the overall result of Cont. et. al. \cite{Cont}.

Furthermore, we use the method of O'Hagen and Leonard \cite{OHagen} to introduce a skew fractional distribution-function that is based on our results in \cite{Jurisch}. The combination of the method of O'Hagen and Leonard with our fractional Gaussian avoids some of the peculiarities that are involved with the L\'evian, and allows us to calculate an acceptable estimate of the local distribution-function that describes the local state of a system.
 
\section{Calculation of L\'evy-parameters}
We start our considerations with the truncated cumulant-function of the fourth order, reading
\begin{equation}
\Psi_{0}(k)\,=\,i\,\sigma_{1}(t)\,k\,-\,\frac{\sigma_{2}(t)}{2}\,k^{2}\,-\,i\frac{\sigma_{3}(t)}{6}k^{3}\,+\,\frac{\sigma_{4}(t)}{24}k^{4}\quad.
\label{levyparameters2}\end{equation}
In our notation we have $\sigma_{1}$ the mean-average, $\sigma_{2}$ the variance, $\sigma_{3}$ the skewness and $\sigma_{4}$ the kurtosis. We denote the cumulants with an argument in time, because they are always to be understood as running cumulants. We can safely assume that a non-Gaussian distribution-function $P(x)$ is sufficiently described by these four cumulants. However, it is impossible to calculate this distribution-function from Eq. (\ref{levyparameters2}) by a Fourier-transform
\begin{equation}
P(x)\,=\,\int_{-\infty}^{\infty}\frac{dk}{2\,\pi}\,\exp[-i\,k\,x\,+\,\Psi_{0}(k)]\quad,
\label{integral}\end{equation}
since the integral will diverge due to the contribution of the kurtosis. The kurtosis but contains interesting information about the shape of the distribution-function, and thus we seek for a method that allows us to include the kurtosis. Note that for $\sigma_{4}=0$ the Fourier-transform Eq. (\ref{integral}) yields an Airy-function that could be used as a skew distribution-function if the oscillating parts are cut off appropriately. We but pursue a more general approach here. The key to solve the problem that is generated by $\Psi_{0}(k)$ is to work with the L\'evian cumulant-function, see e.g. \cite{Feller},
\begin{equation}
\Psi_{\infty}(k)\,=\,\pm\,i\,\gamma\,k\,-\,\alpha_{1}\,|k|^{\alpha_{2}}\,\pm\,i\,\alpha_{1}\,\beta_{2}\,k\,|k|^{\alpha_{2}-1}\,\zeta(\alpha_{2}, k)\quad,
\label{levyparameters11}\end{equation}
where the index $\infty$ denotes the fact that the L\'evian is asymptotically stable. As it will turn out below, the function $\zeta(\alpha_{2}, k)$ will be of no interest here, thus we shall give no further comments on it. The parameter $\gamma$ abbreviates $\gamma=\sigma_{1}-\beta_{1}$, where $\beta_{1}$ is the parameter of interest. The parameter $\beta_{1}$ is not an original L\'evy-parameter, but we have introduced it in order to explain the observations that we have made, see below. Contrary to $\Psi_{\infty}(k)$, the observable cumulant-function $\Psi_{0}(k)$ can be assumed to be valid only for small $k$. 

The problem now is to relate the L\'evy-parameters  $\{\alpha_{1},\,\alpha_{2},\,\beta_{1},\,\beta_{2}\}$ to the cumulants  $\{\sigma_{1},\,\sigma_{2},\,\sigma_{3},\,\sigma_{4}\}$. This can be achieved by a matching with logarithmic conditions, however, the structure of $\Psi_{0}(k)$ is polynomial, and thus does not yield a stable result. A way around this problem is to perform a Pad\'e-resummation on $\Psi_{0}(k)$, as it is done in field-theory in similar situations, see e.g. \cite{Kleinert3}. The Pad\'e-resummed cumulant-function $\mathcal{P}(k)\,=\,\mathcal{P}(k,\,2,\,2)\,+\,i\,\mathcal{P}(k,\,1,\,2)$ is given by
\begin{equation}
\mathcal{P}(k)\,=\,-\,\frac{\sigma_{2}(t)\,|k|^{2}}{2}\,\left(1\,+\,\frac{\sigma_{4}(t)}{12\,\sigma_{2}(t)}|k|^{2}\right)^{-1}\,+\,i\,\sigma_{1}(t)\,k\left(1\,+\,\frac{\sigma_{3}(t)}{6\,\sigma_{1}(t)}\,|k|^{2}\right)^{-1}\quad.
\label{calculation3}\end{equation}
By the Pad\'e-resummation no information has been suppressed or changed in a substantial way, however, the divergency of the Fourier-transform Eq. (\ref{integral}) is removed. For the limits we find
\begin{eqnarray}
\lim_{k\rightarrow 0}\mathcal{P}(k)&=&-\,\frac{\sigma_{2}(t)}{2}\,k^{2}\,+\,i\,\sigma_{1}(t)\,k\quad,\\
\lim_{k\rightarrow\infty}\mathcal{P}(k)&=&-\,6\,\frac{\sigma_{2}(t)^{2}}{\sigma_{4}(t)}\,+\,i\,6\,\frac{\sigma_{1}(t)^{2}}{\sigma_{3}(t)}\,k^{-1}\quad.
\end{eqnarray}
The limits clarify that for small $k$ the characteristic function $\Phi=\exp[\mathcal{P}]$ behaves like a Gaussian, while skew effects enter only asymptotically. The contribution of the kurtosis then but becomes nothing than a factor. As momentum-space and real-space are inverse to each other, we may deduce that the distribution-function in real-space decays like a Gaussian, and shows skew behaviour around the origin. The Pad\'e-resummed cumulant-function Eq. (\ref{calculation3}) would thus already allow us to calculate the distribution-function $P(x)$ by the Fourier-integral Eq. (\ref{integral}). However, we chose to investigate the relation between the cumulants and the L\'evy-parameters. This will provide us with a deeper and more general understanding about the behaviour of a time-series.

Armed with Eq. (\ref{calculation3}) we are now in the position to carry out the matching procedure.

\subsection{Transport-exponent and variance}
 By matching the real parts of $\mathcal{P}(k)$ and $\Psi_{\infty}(k)$ continuously together we obtain an expression for the transport-exponent $\alpha_{2}$,
\begin{equation}
\alpha_{2}(k)\,=\,\frac{24\,\sigma_{2}(t)}{12\,\sigma_{2}(t)\,+\,\sigma_{4}(t)\,k^{2}}\quad.
\label{exponent}\end{equation}
So far, Eq. (\ref{exponent}) depends on the wave-vector. The case $k\,=\,0$ corresponds to $\sigma_{4}(t)\,=\,0$ and gives $\alpha_{2}\,=\,2$, which is the Gaussian. The case of $|k|\,\rightarrow\,\infty$ gives $\alpha_{2}\,=\,0$, which is the extreme fat-tailed regime. A transition region in between may be assumed to be given by the range of the standard-deviation, such that $k_{\rm{match}}\,\approx\,1/\sqrt{\sigma_{2}(t)}$. This assumption is sound, since the standard-deviation approximately defines a region where the crossover from the body of the distribution-function to it's tail takes place. When we insert $k_{\rm{match}}$ into Eq. (\ref{exponent}) we obtain
\begin{equation}
\alpha_{2}\,=\,\frac{24\,\sigma_{2}^{2}(t)}{12\,\sigma_{2}^{2}(t)\,+\,\sigma_{4}(t)}\quad.
\label{exponent1}\end{equation}
By introducing the normalized kurtosis $\tilde{\sigma}_{4}(t)$, we can cast Eq. (\ref{exponent1}) into
\begin{equation}
\alpha_{2}(t)\,=\,\frac{24}{12\,+\,\tilde{\sigma}_{4}(t)},\, \tilde{\sigma}_{4}(t)\,=\,\frac{\sigma_{4}(t)}{\sigma_{2}^{2}(t)}\quad.
\label{exponent2}\end{equation}
This result supports our assumption about $k_{\rm{match}}$, since $\alpha_{2}(t)$ indeed is closely related to the kurtosis of a time-series. Remind that for $\alpha_{2}(t)<2$ the L\'evian is fat-tailed, while for $\alpha_{2}(t)>2$ the L\'evian is thin-tailed.

We have called $\alpha_{2}(t)$ a transport-exponent, because it's numerical value describes the transport-properties that are present in a stochastic system: ordinary diffusion (Gaussian), leptokurtic or jump diffusion for $\tilde{\sigma}_{4}(t)>0$, hence $\alpha_{2}(t)<2$, and platykurtic or Ohmic diffusion for $\tilde{\sigma}_{4}(t)<0$, hence $\alpha_{2}(t)>2$ \footnote{Ohmic diffusion here is meant in the sense that the transport is almost undisturbed by fluctuations.}.

Furthermore, we obtain
\begin{equation}
\alpha_{1}(t)\,=\,\frac{6\,\sigma_{2}(t)^{\alpha_{2}/2}}{12\,+\,\tilde{\sigma}_{4}(t)}\,=\,\frac{\alpha_{2}}{4}\sigma_{2}(t)^{\alpha_{2}/2}\quad,
\label{exponent3}\end{equation}
from which we can define an effective variance
\begin{equation}
\Sigma_{2}(t)\,=\,\frac{\alpha_{2}}{2}\,\sigma_{2}(t)^{\alpha_{2}/2}\quad.
\label{effectivevariance}\end{equation}
By Eqs. (\ref{exponent2}, \ref{exponent3}) the real part of the L\'evian cumulant-function Eq. (\ref{levyparameters11}) is fully determined by the obseravbles $\{\sigma_{2}, \tilde{\sigma}_{4}\}$.

\subsection{Skewness}
The parameters $\{\beta_{1}(t),\,\beta_{2}(t)\}$ describe the skew behaviour of the distribution-function. They can be obtained by matching the imaginary parts of $\mathcal{P}(k)$ and $\Psi_{\infty}(k)$ continuously  as above. For $\alpha_{2}(t)\,\neq\,1$ we find
\begin{equation}
\gamma(t)\,=\,\sigma_{1}(t)\,-\,\beta_{1}(t)\,=\,-\,\frac{72\sigma_{1}^{3}(t)}{(\alpha_{2}(t)-1)\left(6\sigma_{1}(t)+\sqrt{\sigma_{2}(t)}\tilde{\sigma}_{3}(t)\right)^{2}}
+\,\frac{6(1+\alpha_{2}(t))\sigma_{1}^{2}(t)}{(\alpha_{2}(t)-1)\left(6\sigma_{1}(t)+\sqrt{\sigma_{2}(t)}\tilde{\sigma}_{3}(t)\right)}\quad,
\label{skewness1}\end{equation}
and
\begin{equation}
\beta_{2}(t)\,=\,\pm\,\frac{12\sigma_{1}^{2}(t)\sigma_{2}^{\alpha_{2}(t)/2}\tilde{\sigma}_{3}(t)}{\alpha_{1}(t)(\alpha_{2}(t)-1)\left(6\sigma_{1}(t)+\sqrt{\sigma_{2}(t)}\tilde{\sigma}_{3}(t)\right)^{2}\zeta(\alpha_{2},\,k)}\quad.
\label{skewness2}\end{equation}
In Eqs. (\ref{skewness1}, \ref{skewness2}) we have used the normalized skewness
\begin{equation}
\tilde{\sigma}_{3}(t)\,=\,\frac{\sigma_{3}(t)}{\sqrt{\sigma_{2}^{3}(t)}}\quad.
\label{skewness3}\end{equation}
The case of $\alpha_{2}(t)\,=\,1$ is the special case of the Cauchy-distribution, which we shall not discuss here. Our analysis below will show that for stock-market data $\alpha_{2}(t)\,>\,1$ will hold. By inspection of Eq. (\ref{levyparameters11}) we note that the parameter $\beta_{2}(t)$ is multiplied by $\zeta(\alpha_{2},\,k)$, such that we can omit this factor in Eq. (\ref{skewness2}) in our following discussion.

Applied on data, the parameters $\{\gamma(t),\,\beta_{2}(t)\}$ can give large numerical values, which do not match with the behaviour of the time-series under consideration. However, Eqs. (\ref{skewness1}, \ref{skewness2}) are still results of a perturbative approach, and thus we may assume that only the leading order in $\tilde{\sigma}_{3}(t)$ should be taken into account. Moreover, experience shows that the skewness $\tilde{\sigma}_{3}(t)$ usually has only small numerical values compared to $\{\sigma_{1}(t),\,\sigma_{2}(t)\}$, hence the leading order of Eqs. (\ref{skewness1}, \ref{skewness2}) is sufficient anyway. Below we will see that this assumption is indeed correct, since the leading order provides an excellent description of the skew behaviour of a time-series. The leading order in $\tilde{\sigma}_{3}(t)$ of $\{\gamma(t),\,\beta_{2}(t)\}$ is given by
\begin{equation}
\gamma(t)\,=\,\sigma_{1}(t)\,-\,\beta_{1}(t)\,=\,\sigma_{1}(t)\,-\,\frac{(\alpha_{2}(t)-3)\sqrt{\sigma_{2}(t)}}{(\alpha_{2}(t)-1)\,6}\,\tilde{\sigma}_{3}(t)\,+\,\mathcal{O}\left(\tilde{\sigma}_{3}^{2}\right)\quad,
\label{skewness4}\end{equation}
and
\begin{equation}
\beta_{2}\,=\,\pm\,\frac{4\,\tilde{\sigma}_{3}(t)}{3\,(\alpha_{2}(t)-\alpha_{2}(t)^{2})}\,\,+\,\mathcal{O}\left(\tilde{\sigma}_{3}^{2}\right)\quad.
\label{skewness5}\end{equation}
We chose the positive sign in Eq. (\ref{skewness5}), because this choice of the phase coincidences with our observations, see below.

From Eq. (\ref{skewness4}) we can deduce that $\beta_{1}(t)$ describes a skew-shift of the mean-average $\sigma_{1}(t)$, which in the case of a symmetric distribution-function is also it's maximum. A similar phenomenon can be observed in the case of an Airy-distribution with appropriate cut-off, as we already have mentioned above. The parameter $\beta_{2}(t)$, Eq. (\ref{skewness5}), acts like an elasticity, that describes the skew deformation of the distribution-function.

\section{L\'evian statistical mechanics and types of phase-transitions}
In the sense of statistical mechanics the real part of the L\'evian cumulant-function Eq. (\ref{levyparameters11}) has the meaning of a kinetic energy, thus we have the Hamiltonian
\begin{equation}
\mathcal{H}(k)\,=\,\frac{1}{2}\,|k|^{\alpha_{2}}\quad.
\label{thermo1}\end{equation}
Note that the inverse of the effective variance $\Sigma_{2}^{-1}(t)$ in the following plays the role of the thermal energy $k_{\mathrm{B}}\,T$\,. From Eq. (\ref{thermo1}) we immediately can deduce that for $1\leq\alpha_{2}(t)\leq2$ a random-motion varies between classical (diffusive) behaviour for $\alpha_{2}(t)=2$ (Gaussian), and sonic (wave-like) behaviour for $\alpha_{2}(t)=1$\,, which corresponds to the Cauchy-case. This is the first type of a continuous phase-transition that locally occurs if the transport-exponent is a function of time. Concerning the Hamiltonian Eq. (\ref{thermo1}) we shall discuss how e.g. relativistic corrections may lead to fractal behaviour. For low momenta we find
\begin{equation}
\mathcal{H}(p)\,=\,\sqrt{p^{2}\,c^{2}\,+\,m^{2}\,c^{4}}\,=\,m\,c^{2}\,+\,\frac{p^{2}}{2\,m}\,-\,\frac{p^{4}}{8\,m^{3}\,c^{2}}\,+\,\mathcal{O}(p^{6})\quad.
\label{rela1}\end{equation}
By omitting the rest-energy, and by exploiting the relation between the Hamiltonian and the cumulant-function $\Psi_{0}=-\beta\mathcal{H}$, $\beta^{-1}=k_{\mathrm{B}}T$, we find a kurtosis $\tilde{\sigma}_{4}\,=\,3\/(m\,c^{2}\,\beta)$. Note that here $\sigma_{2}=\beta/m$ holds. A statistics that wants to include the relativistic correction suffers from the same problem of convergence as does the Fourier-transform Eq. (\ref{integral}). However, by Eq. (\ref{exponent1}) we find a transport-exponent
\begin{equation}
\alpha_{2}\,=\,2\,-\,\frac{1}{2\,m\,c^{2}\,\beta}\,+\,\mathcal{O}\left(\tilde{\sigma}_{4}^{2}\right)\quad.
\label{rela2}\end{equation}
The inclusion of the first relativistic correction may thus effectively be understood as the introduction of fractal behaviour. For the ultra-relativistic case $\mathcal{H}(p)\,=\,c\,|p|$ we find a Cauchy-distribution. Finally note, that fractal behaviour may also be introduced for a polynomial interaction $U(x)$ up to the fourth order, such that effectively $U(x)\sim x^{\alpha_{2}}$ emerges.

The second type of a continuous phase-transition is related to the first one, or rather a different interpretation of it. The Fourier-transform of the symmetric L\'evian can be cast into
\begin{equation}
P(x)\,=\,\frac{2}{\alpha_{2}}\,\int_{0}^{\infty}\frac{dq}{\pi}\,q^{2/\alpha_{2}-1}\,\exp\left[-\,\frac{1}{2}\,\Sigma_{2}\,q^{2}\right]\,\cos\left[x\,q^{2/\alpha_{2}}\right]\quad,
\label{thermo2}\end{equation}
where we have set $k=q^{2/\alpha_{2}}$. When we interpret Eq. (\ref{thermo2}) as a dimensional integral we find $d(t)=2/\alpha_{2}(t)$ for the dimension. The Gaussian case is given by $d(t)=1$, while for the Cauchy-case it holds $d(t)=2$. Consequently, the continuous phase-transition between classical and sonic motion corresponds to a continuous dimensional phase-transition that dwells between one and two dimensions. For values $\alpha_{2}(t)<1$ the dimension of the system grows rapidly towards infinity. In relation to the tail of the distribution-function we know that large moves become more likely with fatter tails, and this, by $d(t)=2/\alpha_{2}(t)$, corresponds to a growth of the internal degrees of freedom. In the extreme leptokurtic limit $\alpha_{2}(t)\rightarrow 0$ this leads to the analogue of an evaporation in momentum-space. Note that by the continuous variation of $\alpha_{2}(t)$ the distribution-function is in a breathing-mode.

Now we want to discuss the thermodynamic properties of the L\'evian system. Since there is no interaction in real space we may assume that the L\'evian gas is an ideal, albeit fractional gas. Please note, that by our findings above fractal behaviour may very well be related to polynomial interactions. We calculate the partition-function by
\begin{equation}
Z\,=\,\frac{V^{N}}{N!}\,\left(\frac{4}{\alpha_{2}}\,\int_{0}^{\infty}\,dq\,q^{2/\alpha_{2}-1}\,\exp\left[-\,\frac{1}{2}\,\Sigma_{2}\,q^{2}\right]\,\cos\left[\frac{1}{2}\,\Sigma_{2}\,\beta_{2}\,q^{2}\right]\right)^{N}\quad.
\label{thermo3}\end{equation}
In Eq. (\ref{thermo3}) we use a density that is only positive-semidefinite, and thus not a density in a strict sense. The inclusion of the cosine but is the only possibility to include skew behaviour, and we will see that this indeed will make sense. A reader who is uneasy with the cosine here might think of the Wigner-function, which also is only positive-semidefinite, but may be regarded as a phase-space distribution still.

For the free energy, with Stirling's formula, we find
\begin{equation}
F\,=\,-\,\frac{N}{\Sigma_{2}}\,\left(1\,+\,\ln\left[\frac{V}{N\,\alpha_{2}}\,2^{1+1/\alpha_{2}}\Sigma_{2}^{-1/\alpha_{2}}(1+\beta_{2})^{-1/(2\alpha_{2})}\cos\left[\frac{\arctan[|\beta_{2}|]}{\alpha_{2}}\right]\Gamma[1/\alpha_{2}]\right]\right)\quad.
\label{thermo4}\end{equation}

The equation of state follows by
\begin{equation}
-\,\left(\frac{\partial\,F}{\partial\,V}\right)_{\Sigma_{2},\,N}\,=\,p\,=\,\frac{N}{V}\,\Sigma_{2}^{-1}\quad,
\label{thermo5}\end{equation}
where $p$ is the equivalent to the pressure. Eq. (\ref{thermo5}) is the equivalent of the equation of state of an ideal gas, as it was to be expected. We deduce that it holds $p(\alpha_{2})\geq p(\alpha_{2}=2)$, such that the pressure in the non-Gaussian system is always higher than in the Gaussian system. This result is sound, since by the thin body and the fat tail of the L\'evian the probability for large jumps is higher than in the Gaussian case. The higher probability for outbreaks may easily be identified with a higher pressure inside the system.

\begin{figure}[t!]\centering\vspace{-.05cm}
\rotatebox{0.0}{\scalebox{0.87}{\includegraphics{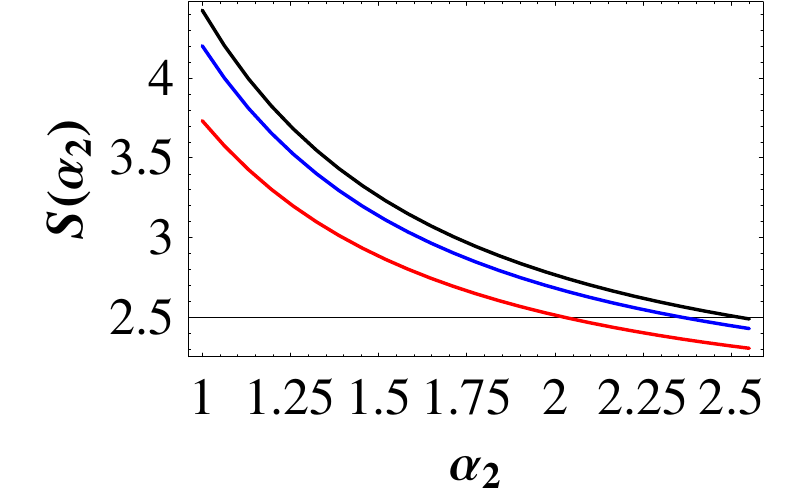}}}
\rotatebox{0.0}{\scalebox{0.91}{\includegraphics{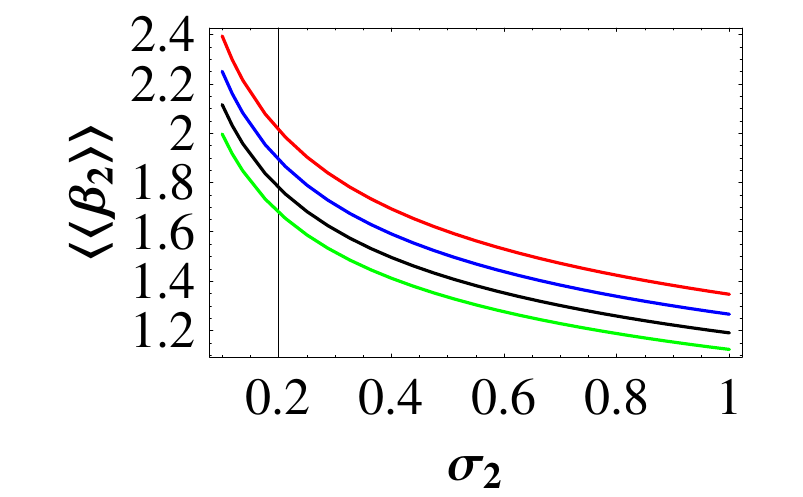}}}
\caption{\footnotesize{Left figure: Entropy as a function of $\alpha_{2}$. Parameters are chosen by $\sigma_{2}=0.5$, $\beta_{2}=0$ (black), $\beta_{2}=0.5$ (blue) and $\beta_{2}=1$ (red). Right figure: Standard-deviation of $\beta_{2}$ as a function of $\sigma_{2}$. Parameters are chosen by $\alpha_{2}=2$ (black), $\alpha_{2}=1.7$ (blue), $\alpha_{2}=1.5$ (red) and $\alpha_{2}=2.5$ (green). General setting is $N=1$, $V=1$\,.}}
\label{Thermodynamics}\end{figure}

For the entropy we find
\begin{equation}
S\,=\,(\Sigma_{2})^{2}\left(\frac{\partial\,F}{\partial\,\Sigma_{2}}\right)_{V,\,N}\,=\,\frac{N\,(1\,+\,\alpha_{2})}{\alpha_{2}}\,+\,N\,\ln\left[\frac{V}{N\,\alpha_{2}}\,2^{1+1/\alpha_{2}}\Sigma_{2}^{-1/\alpha_{2}}(1+\beta_{2})^{-1/(2\alpha_{2})}\cos\left[\frac{\arctan[|\beta_{2}|]}{\alpha_{2}}\right]\Gamma[1/\alpha_{2}]\right]\,.
\label{thermo6}\end{equation}
In Fig. (\ref{Thermodynamics}) we illustrate the entropy $S$ as a function of $\alpha_{2}$ for several values of $\beta_{2}$. Fig. (\ref{Thermodynamics}) elucidates that the elasticity $\beta_{2}$ acts like an order-parameter. The larger the value of $\beta_{2}$, the less entropy is present inside the system. As entropy is a measure for disorder we may conclude that skewness creates order, since the skewness indicates the likely direction of the random-motion. The elasticity, as an order-parameter, may thus be regarded as an analogue to a magnetization or a polarization. By the strict L\'evian condition $-1\leq\beta_{2}\leq1$ the analogue to the ferromagnetic phase-transition is related to $|\beta_{2}|=1$\,.

A further analysis of $\beta_{2}$ in terms of a magnetization can be done by looking at it's moments. We calculate
 \begin{equation}
\left<\beta_{2}^{n}\right>\,=\,\frac{4}{\alpha_{2}}\,\int_{0}^{\infty}\,dq\,q^{2/\alpha_{2}-1}\,\int_{-1}^{1}\,d\beta_{2}\,\beta_{2}^{n}\,\exp\left[-\,\frac{1}{2}\,\Sigma_{2}\,q^{2}\right]\,\cos\left[\frac{1}{2}\,\Sigma_{2}\,\beta_{2}\,q^{2}\right]\quad.
\label{thermo7}\end{equation}
We find that $\left<\beta_{2}\right>=0$\,. This is to be expected, since we may very well assume that the average state of the system is symmetric (paramagnetic), and thus a state of maximum entropy, as we know from above. Note that in the state-analysis of the DAX and the S\&P 500 below we will find $\left<\beta_{2}(t)\right>\neq 0$\,. This result is but still a dynamic average about a limited time-frame, such that there is no principal contradiction to our present thermodynamic result. Only for an ideal infinite time-frame we may expect a stable match with the thermodynamic limit. 

The variance $\left<\beta_{2}^{2}\right>$ is finite. However, we shall not write down the formula since it is cumbrous, the integration but is straight forward. We illustrate the standard-deviation $\left<\left<\beta_{2}\right>\right>=\sqrt{\left<\beta_{2}^{2}\right>}$ in Fig. (\ref{Thermodynamics}). We easily deduce that we can expect skew fluctuations to be stronger in the non-Gaussian case than in the Gaussian case. This result is sound, since the Gaussian case may be regarded as an equilibrium, and an equilibrium can be expected to be symmetric. Additionally, we may deduce that skew fluctuations increase remarkably as $\sigma_{2}$ tends to zero. This state gives a thin body of the corresponding distribution-function, both Gaussian and non-Gaussian, and we certainly may expect an outbreak of the random-motion in this state. An outbreak necessarily but takes place in a certain direction, and the direction of the outbreak then induces a skewness or a polarization of the system. However, we must emphasize that the increase of fluctuations does not necessarily lead to an outbreak. We deal with a random-motion, which very well may relax towards an equilibrium if a highly probable outbreak does not take place, for what reason ever.

\section{State-analysis of the DAX and the S\&P 500}
In this section, we apply our results on the German Dax and on the US-American S\&P 500\,. Our state-analysis will show that the formulas we have derived for the L\'evy-parameters above work well, and allow a satisfactory description of the state a random-motion dwells in.

\subsection{DAX data}
\begin{figure}[t!]\centering\vspace{0.5cm}\hspace{.0cm}
\rotatebox{90.0}{\scalebox{0.42}{\includegraphics{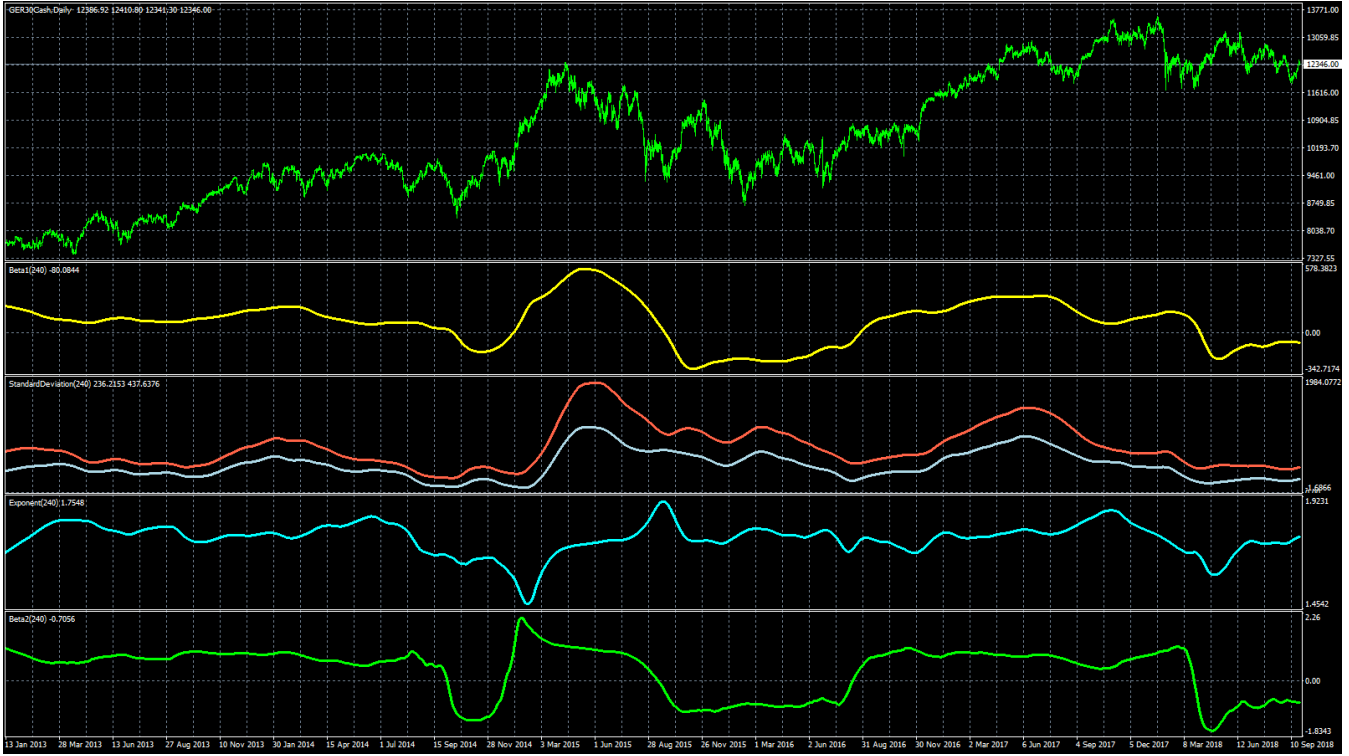}}}
\caption{\footnotesize{Visualization of the DAX-analysis. Windows from top to bottom. Random-motion of the DAX (light-green). Skew-shift parameter $\beta_{1}$ (yellow). Standard-deviations: L\'evian (grey), Gaussian (red). Transport-exponent $\alpha_{2}$ (aqua). Elasticity-parameter $\beta_{2}$ (lime). The running parameters are sampled about a years period, p\,=\,240 (trading days). Data is taken from Metatrader 4, provider XM.com}}
\label{D1picture}\end{figure}
In Fig. (\ref{D1picture}) we illustrate the state-analysis of the German DAX. The time-period is taken from February 2013 to September 2018, the time-unit on the chart is one day. The moving averages are sampled about 240 data points, given that a month has 20 trading-days. 

The windows in the figure from top to bottom are: a) random-motion of the DAX (light-green), b) skew-shift parameter $\beta_{1}(t)$ (yellow), c) standard-deviations, L\'evian $\sqrt{\Sigma_{2}(t)}$ (grey), Gaussian $\sqrt{\sigma_{2}(t)}$ (red), d) transport-exponent $\alpha_{2}(t)$ (aqua), e) elasticity-parameter $\beta_{2}(t)$ (lime). The data is analyzed by the Metatrader software, data is provided by XM.com.

At first we read off that the range of the transport-exponent oscillates between $1.48\leq\alpha_{2}(t)\leq 1.9$\,, the average but is $\left<\alpha_{2}(t)\right>=1.73$\,. This is in perfect agreement with the findings of Cont et. al. \cite{Cont}, where a static overall value of $\alpha_{2}=1.7$ has been found. We also deduce that for a falling $\alpha_{2}(t)$ the DAX performs larger motions, while for a growing $\alpha_{2}(t)$ the motions calms down, and may even stagnate. This also is perfectly in agreement with the fact that the tail of the L\'evian becomes fatter the smaller $\alpha_{2}(t)$ is. Fatter tails indicate that larger jumps become more likely than in a Gaussian or close to Gaussian environment. A close to Gaussian environment may be given for $1.8<\alpha_{2}(t)\leq2$\,. We also but notice that $\alpha_{2}(t)$ declines in times where the fluctuation of the random-motion is restricted to a narrow region. This also is in perfect agreement with the properties of the L\'evian, since for $\alpha_{2}<2$ the body of the distribution-function becomes thin. A fat tail and a thin body are related to each other, such that this state of the random-motion may very well be interpreted in the way that an outbreak is near, as this is a state of high internal pressure. However, it is of course possible that no outbreak takes place, and hence the system relaxes towards the Gaussian regime.

The behaviour of $\alpha_{2}(t)$ is reflected in the behaviour of the standard-deviations. The Gaussian standard-deviation is always larger than the L\'evian standard-deviation. This is to be expected, since for $\alpha_{2}(t)<2$ the effective variance $\Sigma_{2}(t)$, see Eq. (\ref{effectivevariance}), is always smaller than the Gaussian variance $\sigma_{2}(t)$. This again reflects the thin body of the L\'evian and thus the higher pressure.

Furthermore, we find that the skew-shift parameter $\beta_{1}(t)$ indicates a rising motion for $\beta_{1}(t)>0$\,, while for $\beta_{1}(t)<0$ there is a tendency for a contraction or a stagnation. The range of the skewness itself, not plotted in Fig. (\ref{D1picture}), is given by $-1.7\leq\tilde{\sigma}_{3}(t)\leq 1.63$\,. For $\tilde{\sigma}_{3}(t)<0$ we have a rising motion, while for $\tilde{\sigma}_{3}(t)>0$ we have a stagnation or a contraction.

We see that the elasticity $\beta_{2}(t)$ oscillates between $-1.83\leq\beta_{2}(t)\leq 2.26$\,, such that the restriction for the L\'evian $-1\leq\beta_{2}\leq 1$ is violated. We deduce that for $\beta_{2}(t)>0$ there is a rising motion, while for $\beta_{2}(t)<0$ a contraction or a stagnation occurs. The behaviour of $\beta_{2}(t)$ is thus the same as the behaviour of $\beta_{1}(t)$. The average value, however, lies at $\left<\beta_{2}(t)\right>=0.35$\,. Hence, it is possible to calculate a L\'evian for the average value $\left<\beta_{2}(t)\right>=0.35$\,, but not always for a local $\beta_{2}(t)$. How we can deal with this problem in terms of the distribution-function shall be discussed below.

All L\'evy-parameters are functions of time, such that also extrema, turning-points and zeros have to be taken into account in order to deduce a trend, or a change of the trend. Also note, that moving averages are necessarily somewhat retarded, such that the random-motion and the moving averages do not always overlap in their behaviour on the spot. However, the general survey that is provided by $\{\alpha_{2}(t),\,\sqrt{\Sigma_{2}(t)},\,\beta_{1}(t),\,\beta_{2}(t)\}$ is in good agreement with the behaviour of the random-motion. For the construction of the moving averages we have used to calculate the cumulants, please see the appendix.

\subsection{S\&P 500 data}
\begin{figure}[t!]\centering\vspace{0.5cm}\hspace{1.0cm}
\rotatebox{90.0}{\scalebox{0.42}{\includegraphics{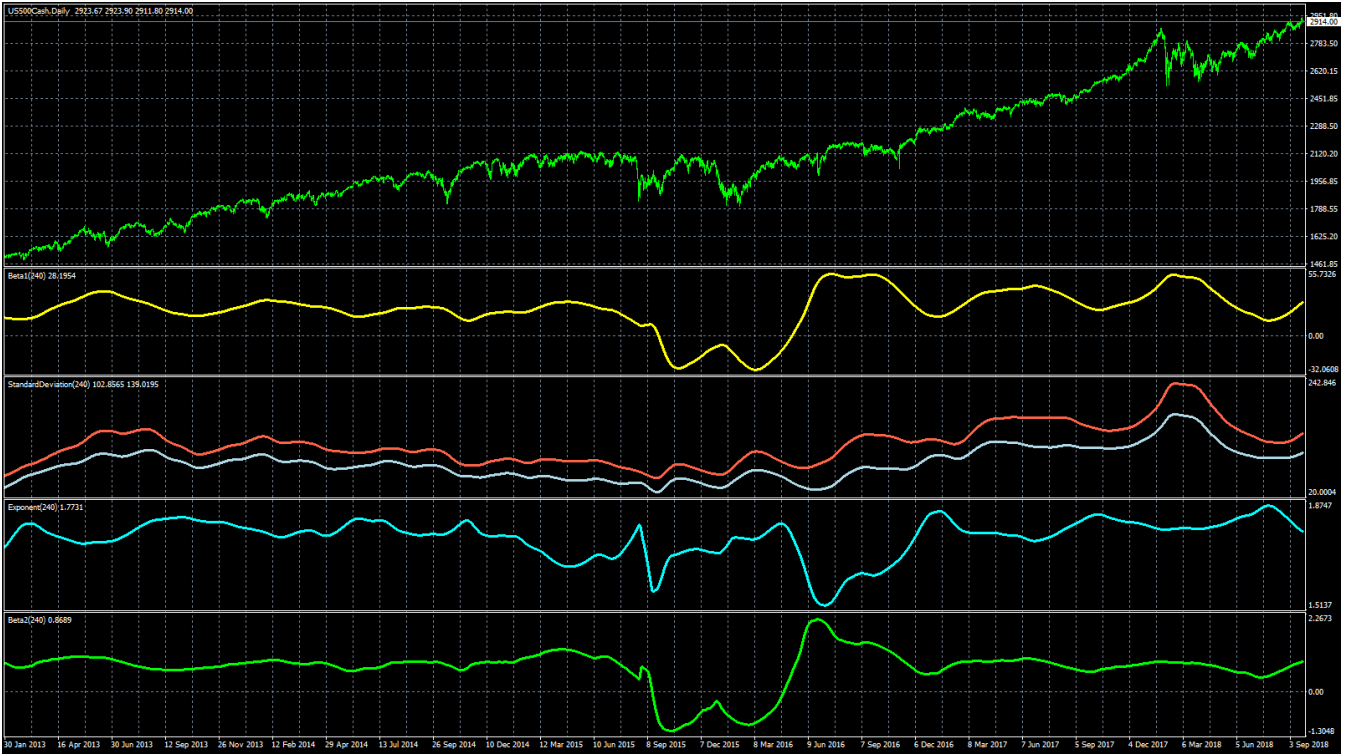}}}
\caption{\footnotesize{Visualization of the S\&P-analysis. Windows from top to bottom. Random-motion of the S\&P 500 (light-green). Skew-shift parameter $\beta_{1}$ (yellow). Standard-deviations: L\'evian (grey), Gaussian (red). Transport-exponent $\alpha_{2}$ (aqua). Elasticity-parameter $\beta_{2}$ (lime). The running parameters are sampled about a years period, p\,=\,240 (trading days). Data is taken from Metatrader 4, provider XM.com}}
\label{SP1picture}\end{figure}

In Fig. (\ref{SP1picture}) we illustrate the state-analysis for the American S\&P 500\,. The time-period is taken from February 2013 to September 2018\,, the time-unit on the chart is one day. The moving averages are again sampled about 240 data points.

The windows in the picture from top to bottom illustrate: a) random-motion of the S\&P 500 (light-green), b) skew-shift parameter $\beta_{1}(t)$ (yellow), c) standard-deviations, L\'evian $\sqrt{\Sigma_{2}(t)}$ (grey), Gaussian $\sqrt{\sigma_{2}(t)}$ (red), d) transport-exponent $\alpha_{2}(t)$ (aqua), e) elasticity-parameter $\beta_{2}(t)$ (lime). The data is analyzed by the Metatrader software, data is provided by XM.com.

At first we notice that the random-motion of the S\&P 500 is calmer than the random-motion of the DAX. This is due to the fact that European indices are weighted performance-indices, while Anglo-Saxon indices are mostly weighted mean-value indices. The DAX thus does not only include the value of it's components but also their turnover, compared to a date of reference in the previous year.

We find a transport-exponent that lies between $1.5\leq\alpha_{2}(t)\leq1.87$\,, the average is given by $\left<\alpha_{2}(t)\right>= 1.74$\,. Again we find perfect agreement with the static overall value of Cont et. al. \cite{Cont}, $\alpha_{2}=1.7$\,. All what we have said above about possible outbreaks or the relaxation towards the Gaussian regime also holds here, of course.

The elasticity lies between $-1.3<\beta_{2}(t)<2.26$\,, and thus again violates the strict L\'evian condition $-1\leq\beta_{2}\leq1$\,. The average of the elasticity, however, is given by $\left<\beta_{2}(t)\right>=0.62$\,, such that an overall L\'evy-distribution is calculable. The range of the skewness itself, not plotted in Fig. (\ref{SP1picture}), is thereby given by $-1.86<\sigma_{3}(t)<1.49$\,.

All what is said above about the other parameters $\{\sqrt{\Sigma_{2}(t)}, \sqrt{\sigma_{2}(t)}, \beta_{1}(t)\}$ also holds here.

\section{Construction of skew distribution-functions}
From our data-analysis above we know, that the L\'evian condition for the elasticity $-1\leq\beta_{2}\leq1$ is violated if it is a local function $\beta_{2}(t)$. However, for a stable L\'evian it unconditionally must hold that $-1\leq\beta_{2}\leq1$\,. This makes the L\'evian in it's strict sense useless for the calculation of the local distribution-function. A way to solve this problem in an elegant way was suggested by O'Hagen and Leonard \cite{OHagen}. In order to discuss how we can still make use of the L\'evy-parameters, we shall give an elementary motivation for this method.

The introduction of the skewness is similar to an anti-symmetrization of a symmetric function. As such, any anti-symmetric function qualifies as a tool to introduce skewness, at least in principal. A general assumption for an anti-symmetrizer is provided by the step-function
\begin{equation}
A(x)\,=\,\theta(x)\quad.
\label{skew1}\end{equation}
The function $A(x)$ as it is given by Eq. (\ref{skew1}) is somewhat a hard anti-symmetrizer.
However, the function $A(x)$ but has the remarkable property that it can be smeared out into cumulative probability-distributions, CDF, e.g.
\begin{equation}
A(x; \beta_{2})\,=\,\frac{1}{2}\left(1+{\rm{erf}}\left[\frac{\beta_{2}\,x}{\sqrt{2\,\sigma_{2}}}\right]\right)\quad,
\label{skew3}\end{equation}
which is related to the CDF of the Gaussian.
For any function that qualifies as a function $A(x; \beta_{2})$ the following limits must hold,
\begin{equation}
\lim_{\beta_{2}\rightarrow 0}A(x; \beta_{2})\,=\,\frac{1}{2}\,,\quad \lim_{\beta_{2}\rightarrow\pm\infty}A(x; \beta_{2})\,=\,\theta(\pm\,x)\quad.
\label{skew4}\end{equation}
The properties that are required by Eq. (\ref{skew4}) are fulfilled by any CDF of a symmetric distribution-function, given a proper normalization.

The construction that is suggested in \cite{OHagen} now is that skewness can be introduced by anti-symmetrizing a symmetric distribution-function by the ansatz
\begin{equation}
P(x; \beta_{2})\,=\,P_{{\rm{sym}}}(x)\,A(x; \beta_{2})\quad.
\label{skew5}\end{equation}
With a proper normalization, the skew distribution-function then obeys the following properties
\begin{equation}
\lim_{\beta_{2}\rightarrow 0}P(x; \beta_{2})\,=\,P_{{\rm{sym}}}(x)\,,\quad \lim_{\beta_{2}\rightarrow\pm\infty}P(x; \beta_{2})\,=\,P_{{\rm{sym}}}(x)\,\theta(\pm x)\quad.
\label{skew6}\end{equation}
The parameter $\beta_{2}$ acts like an elasticity as above. In \cite{OHagen} the skew parameter is estimated by fitting. Armed with our results about the L\'evy-parameters but we are able to circumvent this unhandy approach, and use $\{\beta_{1},\,\beta_{2}\}$ as calculated from the time-series. Consequently, a skew distribution-function with L\'evy-parameters can be proposed by the ansatz
\begin{equation}
P(x; \alpha_{1}, \alpha_{2}, \gamma, \beta_{2})\,=\,P_{{\rm{sym}}}(x-\gamma; \alpha_{1}, \alpha_{2})\,A\left(\beta_{2}(x-\gamma); \alpha_{1}, \alpha_{2}\right)\quad.
\label{skew9}\end{equation}

\begin{figure}[t!]\centering\vspace{0cm}
\rotatebox{0.0}{\scalebox{0.87}{\includegraphics{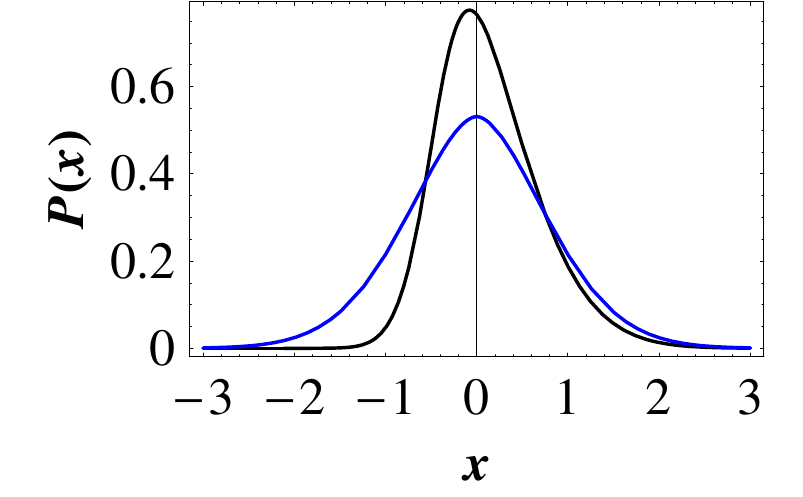}}}
\rotatebox{0.0}{\scalebox{0.91}{\includegraphics{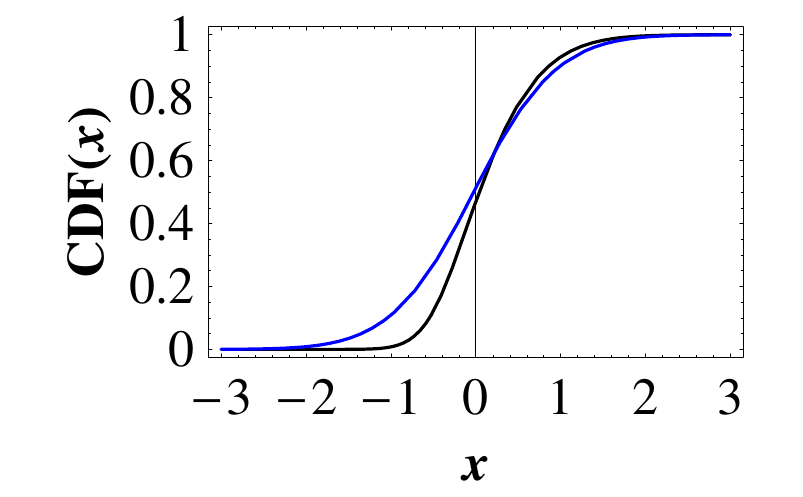}}}
\caption{\footnotesize{Illustration of the skew behaviour according to the anti-symmetrization as given by Eq. (\ref{skew9}). Left:  skew distribution-function Eq. (\ref{skew9}) (black) compared to the symmetric distribution-function Eq. (\ref{fractionalGaussian1}) (blue). The distributions are normalized. Right: illustration of the corresponding cumulative probability-distributions CDF. We have chosen $\alpha_{2}=1.7$, $\sigma_{2}=0.5$ and $\tilde{\sigma}_{3}=-2$.}}
\label{Skew10}\end{figure}
One must always bear in mind that the construction Eq. (\ref{skew9}) is an estimate, and not an exact quantity. A suitable anti-symmetrizer $A$ is always provided by the CDF of $P_{{\rm{sym}}}$.

As a model for $P_{{\rm{sym}}}$ we use the fractional and extremal Gaussian we have derived in \cite{Jurisch}, reading
\begin{equation}
P_{\rm{sym}}(x)
\,=\,N(\alpha_{1},\,\alpha_{2})\,\exp\left[-|x|\left(\frac{|x|}{\alpha_{1}(4-\alpha_{2})}\right)^{\frac{1}{3-\alpha_{2}}}\right]\,
\exp\left[\left(\frac{|x|}{\alpha_{1}(4-\alpha_{2})}\right)^{\frac{4-\alpha_{2}}{3-\alpha_{2}}}\alpha_{1}\right]\quad,
\label{fractionalGaussian1}\end{equation} 
with the normalization
\begin{equation}
N(\alpha_{1},\,\alpha_{2})\,=\,\frac{1}{2}
\left(\frac{3-\alpha_{2}}{4-\alpha_{2}}\left(\frac{1}{\alpha_{1}(4-\alpha_{2})}\right)^{\frac{1}{3-\alpha_{2}}}\right)^{\frac{3-\alpha_{2}}{4-\alpha_{2}}}\,\left(\Gamma\left[\frac{2\,\alpha_{2}-7}{\alpha_{2}-4}\right]\right)^{-1}\quad.
\label{fractionalGaussian2}\end{equation}
For $\alpha_{2}=2$\,, the fractional Gaussian is the normal Gaussian, of course. Our distribution-function has the advantage that we can get rid of the Fourier-transform that has to be carried out in order to obtain a L\'evian or a truncated L\'evian in real space.

As Eq. (\ref{fractionalGaussian1})  already depends on two L\'evy-parameters $\{\alpha_{1},\,\alpha_{2}\}$, it remains to introduce the skew parameters $\{\gamma,\,\beta_{2}\}$ by the construction given by Eq. (\ref{skew9}). The distribution-function Eq. (\ref{fractionalGaussian1}) still decays exponentially, however, this is compensated by a fatter body for $\alpha_{2}<2$\,, such that the space for fluctuations is still increased in comparison to the Gaussian, see \cite{Jurisch} for more insight.

In Fig. (\ref{Skew10}) we illustrate the effect of the anti-symmetrization as discussed above. We clearly see that a reasonable skew distribution-function is generated, that fulfills all requirements that are to be expected. For a negative skewness $\tilde{\sigma}_{3}<0$\,, we observe that the maximum of the skew distribution-function has slightly moved to the left, due to the effect of $\gamma\,=-\,\beta_{1}$. Note that $\sigma_{1}=0$\,. The elasticity $\beta_{2}$ has reduced the space for fluctuations on the left from $\gamma$, while on the right from $\gamma$ the space for fluctuations is slightly increased. This behaviour is right what we would expect from a skew distribution-function.

Consequently, we are now in the position to calculate the distribution-function for any possible value of the L\'evy-parameters, especially the elasticity $\beta_{2}$, free of the restriction that is imposed by the hard constraint $-1\leq\beta_{2}\leq1$\,.

\section{Visualization of the distribution-function on data}
We now can carry the state-analysis of a random-motion one step further. The method discussed above allows us a direct visualization of the distribution-function as a local function of the time-dependent L\'evy-parameters. As examples we again chose the DAX and the S\&P 500\,, but on a larger scale as above. We shall analyze the random-motion on the time-unit of a week. This scale allows us to describe the general state of the random-motion. The sampling-period we have chosen is 260 weeks, thus five years. The visualized time-period is taken from 2015 to 2018\,. Again we have used Metatrader, data provided by XM.com.

\begin{figure}[t!]\centering\vspace{0.5cm}\hspace{0.0cm}
\rotatebox{0.0}{\scalebox{0.6}{\includegraphics{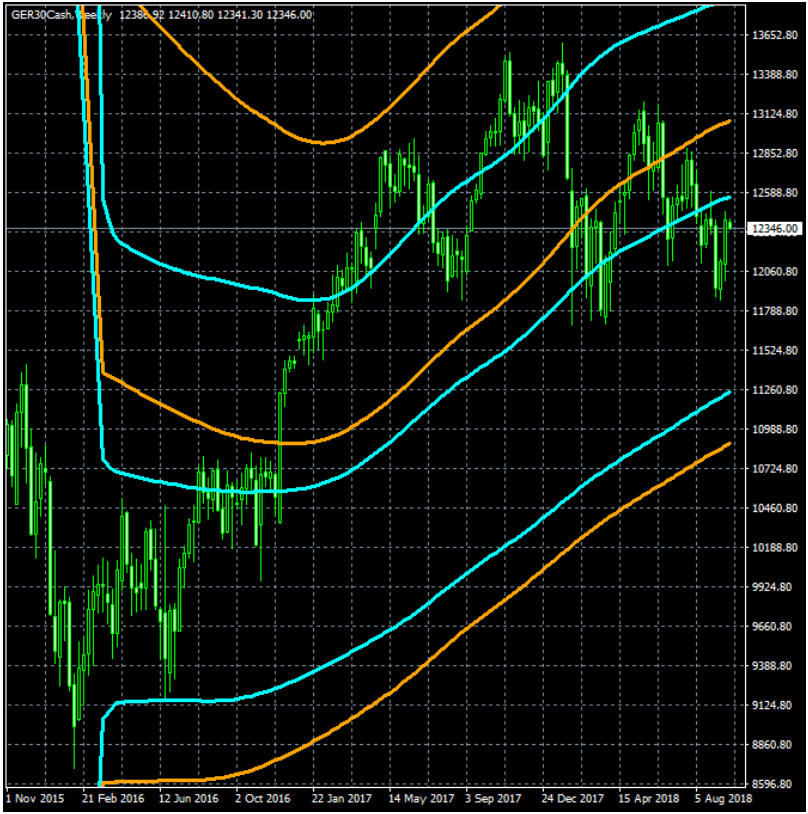}}}
\caption{\footnotesize{Visualization of the DAX-analysis. The running parameters are sampled about a years period, p\,=\,260 (trading weeks). The orange lines visualize the Gaussian, the aqua lines visualize the skew fractional distribution Eq. (\ref{skew9}). Data is taken from Metatrader 4, provider is XM.com}}
\label{D2picture}\end{figure}

\begin{figure}[t!]\centering\vspace{0.5cm}\hspace{0.0cm}
\rotatebox{0.0}{\scalebox{0.6}{\includegraphics{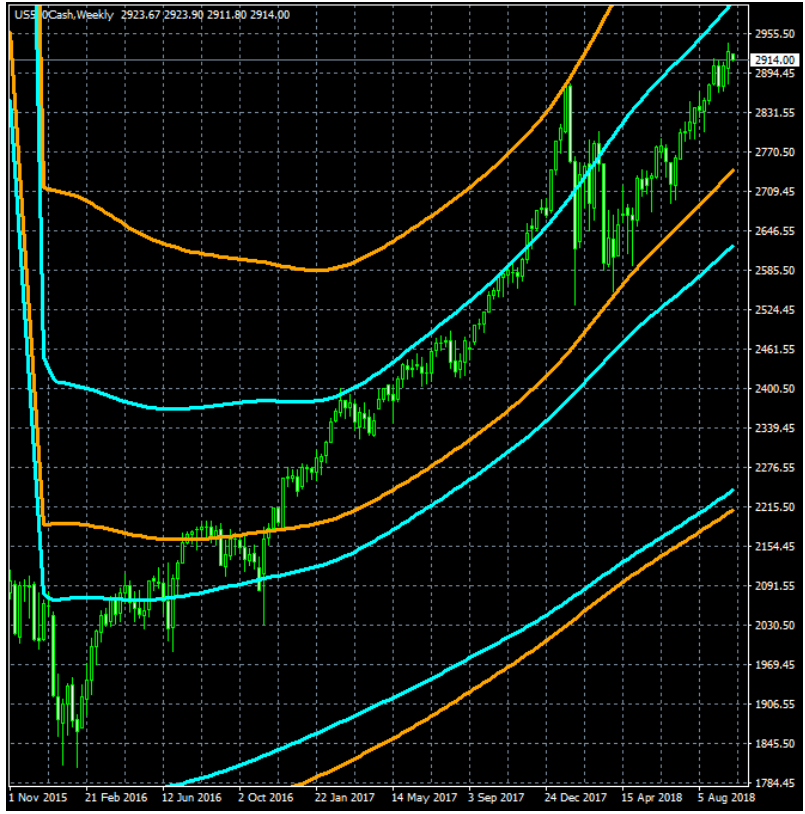}}}
\caption{\footnotesize{Visualization of the S\&P-analysis. The running parameters are sampled about a years period, p\,=\,260 (trading weeks). The orange lines visualize the Gaussian, the aqua lines visualize the skew fractional distribution. Data is taken from Metatrader 4, provider is XM.com}}
\label{SP2picture}\end{figure}

In Figs. (\ref{D2picture}, \ref{SP2picture}) we illustrate the mapped distribution-functions for the DAX and the S\&P 500\,. The orange lines denote the Gaussian, the aqua lines denote the skew fractional distribution. The middle-line is the maximum, while the outer lines mark the standard-deviation. The mapping is done as follows. In the case of the Gaussian we plot
\begin{eqnarray}
\mathrm{Maximum}&:& \sigma_{1}(t)\quad,\nonumber\\
\mathrm{Standard-deviation}&:& \sigma_{1}(t)\,\pm\,\sqrt{\sigma_{2}(t)}\quad.\nonumber
\end{eqnarray}
In the case of the skew fractional distribution we plot
\begin{eqnarray}
\mathrm{Maximum}&:& \gamma(t)\,=\,\sigma_{1}(t)\,-\,\beta_{1}(t)\quad,\nonumber\\
\mathrm{Standard-deviation}&:& \gamma(t)\,\pm\,2\,\sqrt{\Sigma_{2}(t)}\,\mathrm{CDF}\left[\gamma(t)\,\beta_{2}(t)\right]\quad.\nonumber
\end{eqnarray}
By this construction, we fulfill the requirements of the anti-symmetrization as discussed in Eq. (\ref{skew6}). For $\beta_{2}(t)\rightarrow 0$ we recover the symmetric case, while for $\beta_{2}(t)\rightarrow\pm\infty$ we obtain the total anti-symmetric case. Note that the symmetric case is not necessarily the Gaussian regime, since for the transport-exponent it may hold $\alpha_{2}(t)<2$ still.

From Fig. (\ref{D2picture}) we easily can deduce that the random-motion of the DAX shows exactly the skew and fractional behaviour that is to be expected by our knowledge about the behaviour of the L\'evy-parameters. 

First of all, we notice that the skew fractional distribution-function indicates a rising motion in principal, since $\sigma_{1}(t)>\gamma(t)$, which indicates that $\beta_{1}(t)>0$. The lower fluctuation-space (aqua) is slightly reduced, while the upper fluctuation-space (aqua) is slightly enhanced. This behaviour is the same as it is illustrated in Fig. (\ref{Skew10}). In 2015 and 2016 we notice that the fluctuations stay inside the fluctuation-space that is denoted by the aqua lines. Note especially, that the stagnation in 2016 fluctuates exactly around $\gamma(t)$. In 2017 and 2018 we see a crossing of the upper standard-deviation $\Sigma_{2}(t)$ from below, which can be interpreted as an exaggeration. In 2018 then the uptrend is broken by the crossover of $\gamma(t)$ from above, as far as it is plotted.

The same analysis holds for the S\&P 500\,, Fig. (\ref{SP2picture}). As above, we notice that the random-motion of the S\&P 500 is calmer than the DAX. The stagnation in 2015 and 2016 oscillates around $\gamma(t)$, and also inside the lower fluctuation-space that is denoted by the aqua lines. In the end of 2016 an uptrend starts, and at the beginning of 2018 we find an exaggeration of this uptrend. After some turbulences the uptrend resumes, as far as it is plotted.

In summary, we see that our skew fractional description of a random motion is in good agreement with it's actual behaviour. Locally, we find that our approach provides a much better means for interpretation than the Gaussian.

\section{Conclusion}
We have developed a method that allows us to relate the observable properties of a time-series, the cumulants $\{\sigma_{1}, \sigma_{2}, \sigma_{3}, \sigma_{4}\}$, to the L\'evy-parameters $\{\alpha_{1}, \alpha_{2}, \beta_{1}, \beta_{2}\}$. Our method relies on the Pad\'e-resummation of the truncated cumulant-function, Eq. (\ref{levyparameters2}), which allows us to heal the divergency of the Fourier-transform, that relates the characteristic function to the distribution-function. Thus, asymptotic stability is gained, which makes it possible to match the Pad\'e-resummed cumulant-function, Eq. (\ref{calculation3}), continuously on the L\'evian cumulant-function, Eq. (\ref{levyparameters11}).

From the matching we have calculated explicit formulas for the L\'evy-parameters. Our formulas allow us to analyze the state of a random-motion with respect to their non-Gaussian properties. We emphasize that this analysis can be carried out in real-time, since the L\'evy-parameters are time-dependent functions of the observable cumulants. Consequently, a static matching by maximum-likelihood or least-square methods drops away. This removes a considerable amount of effort in the state-analysis of random-motions.

We analyzed the statistical and thermodynamic properties of the L\'evian in terms of a fractional gas. Furthermore, we demonstrated how our formalism easily allows to relate relativistic corrections to the classical kinetic energy to fractal behaviour. The major result of our analysis, however, is that the elasticity $\beta_{2}(t)$ acts like an order-parameter, similar to a magnetization or a polarization. Furthermore, we elucidated the nature of the continuous phase-transitions that take place in a L\'evian system, if it's parameters, and especially the transport-exponent $\alpha_{2}$, are functions of time. Note that the same kind of phase-transitions occur if $\alpha_{2}$ is a function of the temperature.

As examples for the application of our results we have chosen the German DAX and the American S\&P 500\,. A comparison between the random-motion and our formulas shows good agreement in the description of the non-Gaussian properties. This holds especially for the value of the transport-exponent $\alpha_{2}(t)$. The mean-values of the time-dependent exponent are $\left<\alpha_{2}(t)\right>=1.73$ for the DAX, and $\left<\alpha_{2}(t)\right>=1.74$ for the S\&P 500\,. This matches with the results of Cont et. al. \cite{Cont}, where an overall static value of $\alpha_{2}=1.7$ has been found.

Since the local values of the time-dependent elasticity $\beta_{2}(t)$ violate the strict L\'evian condition $-1\leq\beta_{2}\leq 1$\,, it is impossible to calculate a local L\'evian distribution-function. We circumvented this problem by using the extremal and fractional Gaussian we have derived in \cite{Jurisch}. The inclusion of the skewness is done by the construction of O'Hagen and Leonard \cite{OHagen}. The result is a skew distribution-function, that describes the properties of a random-motion in a reasonable way. This is confirmed by a direct mapping of the distribution-function on the random-motion of the indices, as it is illustrated in Figs. (\ref{D2picture}, \ref{SP2picture}).

Further work is given by the question of how our results could be related to the Tsallis-distribution, and thus to non-extensive thermodynamics. This is of particular interest, since a comparison between the L\'evian and Tsallis' approach may elucidate similarities, but certainly also differences between these two major fractional statistics. Remind that there is no unique way to introduce non-Gaussian behaviour.

\begin{appendix}

\section{Weighted moving average} 
The heart of any time-series analysis is the moving average. There are extended methods for the calculation of moving averages in terms of ARMA, ARCH and GARCH. To our regards, all these ans\"atze do not serve the needs of an easy application. Furthermore, as we deal with stochastic systems where the knowledge is limited anyway, we think that it is better to work with a minimum of basic assumptions. Instead, we shall use a weighted moving average that is calculated without any need to determine additional parameters. The moving average of our choice is linear-weighted, given by
\begin{equation}
\left<m(t_{0}, p)\right>\,=\,\sum_{n=0}^{N-1}\,w(n)\,c(t_{n})\quad.
\label{ma1}\end{equation}
The weights $w(n)$ and their normalization are given by
\begin{equation}
 w(n)\,=\,(N-1-n)\left(\sum_{n=0}^{N-1}\,(N-1-n)\right)^{-1}\,
=\,\frac{2\,(N-1-n)}{(N-1)\,N}\quad.
\label{ma2}\end{equation}
In Eq. (\ref{ma2}) $c(t_{n})$ is the closing price at time $t_{n}$. The expression $\left<m(t_{0}, p)\right>$ is the mean-value at time $t_{0}$, calculated for a period $p$. Thereby $p=N$ data-points. The weights $w(n)$ define a triangular decreasing filter-window, such that the newest event is weighted higher than older events. Technically spoken, a triangular window is a low-pass filter. A low-pass filter damps all higher modes, while low modes can pass through the filter almost without a loss of information. This is important, since low-lying modes become more important the smaller the numerical value of $\alpha_{2}$ is. This fact can easily be deduced by the Fourier-transform
\begin{equation}
P(x)\,=\,\int_{-\infty}^{\infty}\frac{dk}{2\,\pi}\,\exp[-\,i\,k\,x\,-\,\alpha_{1}\,|k|^{\alpha_{2}}]\quad.
\end{equation}
The smaller the numerical value of $\alpha_{2}$, the fatter is the body of the characteristic function $\Phi(k)=\exp[-\,\alpha_{1}\,|k|^{\alpha_{2}}]$. A fat body in momentum-space but indicates a fat tail of $P(x)$ in real-space - and vice-verse.

Some further technical remarks shall be in order. The sum in Eq. (\ref{ma1}) is carried out in the reverse direction then it is usually done, because it is common that in the analysis of charts the youngest event is set on the point $t_{0}$. For the analysis of other systems the necessary changes are straight-forward.

\section{Construction of a basic trend-function}
An advantageous slope-line for the moving average is given by the construction \footnote{This construction is widely used in the trader-scene. However, we are not able to provide a clear reference for this construction. It can be found in the Internet by the term \emph{slope-direction line}, used by several authors of technical indicators. The mathematical analysis of Eqs. (\ref{t1}, \ref{t2}), as far as we know, has not yet been carried out by another author.},
\begin{equation}
\left<M(t_{0}, p)\right>\,=\,2\,\left<m(t_{0}, p/2)\right>\,-\,\left<m(t_{0}, p)\right>,\quad\,p\,=\,N\quad.
\label{t1}\end{equation}
The ansatz Eq. (\ref{t1}) doubly overweights the younger half of the $N$-period array, from which the whole array is subtracted. By this a memory-kernel is created, that is less inert than the moving average given by Eq. (\ref{ma2}). Inertness is a major problem in time-series analysis. Moving averages that are too inert are as useless as ones that are not inert enough. 

A further smoothing can be achieved by summing up $\left<M(t_{0}, p)\right>$. This finally gives us our trend function
\begin{equation}
\left<\left<M(t_{0}, p)\right>\right>\,=\,\frac{1}{p}\sum_{n=0}^{p}\,\left<M(t_{n}, p)\right>,\quad p\,=\,\sqrt{N}\quad.
\label{t2}\end{equation}
The square-root in Eq. (\ref{t2}) corresponds to a Gaussian, and effectively removes a considerable part of remaining noise in the weighted moving average. A background reason for this choice can be deduced from fractional calculus. It holds that $\left<x(t)\right>\sim\sqrt{t}$. Please see the exhaustive report of Metzler et. al. \cite{Metzler} for more insight. Numerically, however, it is of course not possible to sum up to the true value of the square-root, it is sufficient to take the closest integer instead.

\subsection{Positive definiteness of the memory-kernel}
Here we shall treat the question why the ansatz for the memory kernel Eq. (\ref{t1}) works. The memory-kernel Eq. (\ref{t1}) can be written by
\begin{equation}
\left<M(t_{0}, p)\right>\,=\,2\sum_{n\,=\,p/2\,-\,1}^{0}\,w(n/2)\,c(t_{n})\,-\,\sum_{n\,=\,p\,-\,1}^{0}\,w(n)\,c(t_{n})\quad,
\label{proof1}\end{equation}
A reordering of the sum in Eq. (\ref{proof1}) leads to
\begin{eqnarray}
2\sum_{n\,=\,p/2\,-1}^{0}a_{n}\,-\sum_{n\,=\,p\,-\,1}^{0}b_{n}\,&=&\,\sum_{n\,=\,p/2\,-\,1}^{0}2a_{n}-b_{n}-b_{n+p/2}\nonumber\\&=&\,\sum_{n\,=\,p/2\,-\,1}^{0}2a_{n}\left(1-\frac{b_{n}+b_{n+p/2}}{2a_{n}}\right)\quad.
\label{proof2}\end{eqnarray}
The criterion we are looking for is whether the bracket in Eq. (\ref{proof2}) is always positive. Inserting the coefficients, we obtain
\begin{equation}
1-\frac{b_{n}+b_{n+p/2}}{2a_{n}}\,=\,1\,-\,\frac{w(n)}{2\,w(n/2)}\,-\,\frac{w(n+p/2)}{2\,w(n/2)}\frac{c(t_{n+p/2})}{c(t_{n})}\quad.
\label{proof3}\end{equation}
For Eq. (\ref{proof3}) we make the assumption that the elements of the dataset are all of the same order of magnitude. Consequently, we may set $c(t_{n+p/2})/c(t_{n})\,\approx\,1$\,. Furthermore, we can deduce that $w(n)/w(n/2)\,<\,1\, \forall\,n$. This finally ensures that Eq. (\ref{proof3}) has positive definiteness for all periods $p$. Thus, the general formula of a possible memory kernel must read
\begin{equation}
\left<M(t_{0}, p)\right>_{\mu}\,=\,\frac{2}{\mu(\mu+1)\,-\,4}\,\left(\sum_{\nu\,=\,2}^{\mu}\,\nu\,\left<m(t_{0}, p/\nu)\right>\,-\,\left<m(t_{0}, p)\right>\right)\quad,
\label{proof4}\end{equation}
where $\mu\,\leq\,p$. Experience shows that $\mu=2$ is enough. For values $\mu>2$ too much inertness is removed.

\subsection{Moving cumulants}
In order to calculate the L\'evy-parameters we also need the moving cumulants. The calculation is  done by
\begin{equation}
\left<C_{k}(t_{0}, p)\right>\,=\,\sum_{n=0}^{N-1}w(n)\left(c(t_{n})\,-\,\left<M(t_{0}, p)\right>\right)^{k}\,,\quad k\,=\,2\,,3\,,4\quad.
\label{cu1}\end{equation}
The cumulants $\left<C_{k}(t_{0}, p)\right>$ are then further smoothed by the procedure described by Eq. (\ref{t2}),
\begin{equation}
\sigma_{k}(t_{0}, p)\, =\,\frac{1}{p}\sum_{n=0}^{p}\,\left<C(t_{n}, p)\right>,\quad p\,=\,\sqrt{N}\quad.
\label{cu2}\end{equation}
The smoothed cumulants are used to calculate the L\'evy-parameters.

\end{appendix}

%\end{mainmatter}

\end{document}